\documentclass{evn2004}
\usepackage{txfonts}
\usepackage{graphicx}
\begin{document}
\setcounter{page}{235}

   \title{Microquasar Observations with MERLIN}

   \author{I.~K. Brown}

   \institute{Jodrell Bank Observatory, University of Manchester}

   \abstract{  A series of observations of microquasar  GRS1915+105 were made 
   by  the MERLIN telescope  array  in  the   spring  of  2003 at 18 cm.  The images show polarization in the jet components, with two epochs displaying Faraday rotation of  $\sim$ 2500 rad/$m^2$. This effect is a demonstration of the similarities between microquasars and extragalactic jets.

   }

   \maketitle
%

\section{Introduction}

Of  the X-ray  binaries in  the galaxy,  about 50 are also detectable  at radio
wavelengths. This is thought to  be synchrotron radiation from relativistic
jets. Their similarity to quasars except for much smaller scales and 
distances from us has named these sources microquasars. Effects such as 
apparent superluminal motion are now seen in both and here we look for 
another feature seen in extragalactic jets, Faraday rotation, 
the rotation of the plane of polarization with respect to wavelength.

One object considered a microquasar is GRS1915+105. It was first found
in  X-rays   with  the  watch   instrument  on  the   GRANAT  satellite
(Castro-Tirado  et al. \cite{Castro-Tirado}) before a radio
counterpart was found  (Mirabel et al \cite{Mirabel}).  It is  best known as the first source in our galaxy to show apparent superluminal motion (Mirabel \&  Rodr{\' i}guez \cite{Mirabela}) . These
ejections have allowed an estimate to the upper limit to its distance of
$11.2\pm0.8$ kpc (Fender et al \cite{Fender}).


\section{Observations}
The times of the 5 observations  of GRS1915+105 by MERLIN are in table
1. All the observations were made at 1658.0 MHz with 16 MHz bandwidth. Also
observed was a phase calibrator  1919+086, a flux calibrator 3C286 and
point  source   0552+368.   The   data sets  from  the   observations  were
transformed into  FITS files by telescope specific  programs at Jodrell
Bank Observatory that also set  the flux scale from 3C286.   The FITS 
files where  then loaded into NRAO AIPS.  bad data was
removed using the  AIPS task IBLED. CALIB and CLCAL  were then used to
calibrate  1919+086. The  gain corrections  were then  applied  to the
target  source. To  correct for  the leakage  of flux  from  one cross
polarized mode  to the other, the  task PCAL was run  on 1919+086. The
polarization angle was then calibrated with respect to 3C286. Finally 
the target source went through several loops of self calibration.

To measure the Faraday rotation each epoch was split into four subbands
and images were made in total intensity, polarized intensity and polarization angle. The position of the peak 
polarized intensity of a component was used when measuring the angle.

   \begin{table}
      \caption[]{Times of MERLIN observations of GRS1915+105 in 2003 }
         \label{times}
         \begin{tabular}{p{0.5\linewidth}p{0.5\linewidth}}
            \hline
            \noalign{\smallskip}
            Start Time      &  End Time \\
            \noalign{\smallskip}
            \hline
            \noalign{\smallskip}
	    March 06 02:00 & March 06 14:40 \\
	    March 24 09:33 & March 25 13:14 \\
            April 17 23:45 & April 18 10:45 \\
	    May 09 22:00 &  May 10 10:00 \\
	    June 15 19:30 & June 16 07:45 \\
            \noalign{\smallskip}
            \hline
         \end{tabular} 
   \end{table}
%

   \begin{figure}
   \centering
   \includegraphics[angle=-90, width=7.5cm]{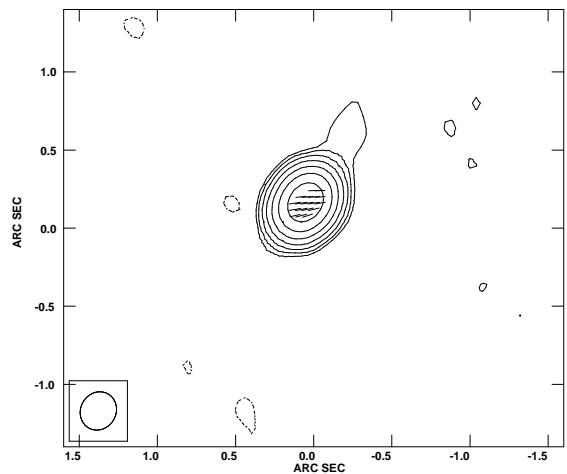}
   \caption{GRS1915+105 March 6. The contours go up in powers of two from 0.24 mJy/beam, the bars give polarized intensity and angle, 1 arcsec equals 12.5 mJy /beam. observed PA at 1658 MHz is $-81.3^{o}$, 3\% polarization.
            \label{fig:march}
         }
   \end{figure}
%

   \begin{figure}
   \centering
   \includegraphics[width=7.5cm]{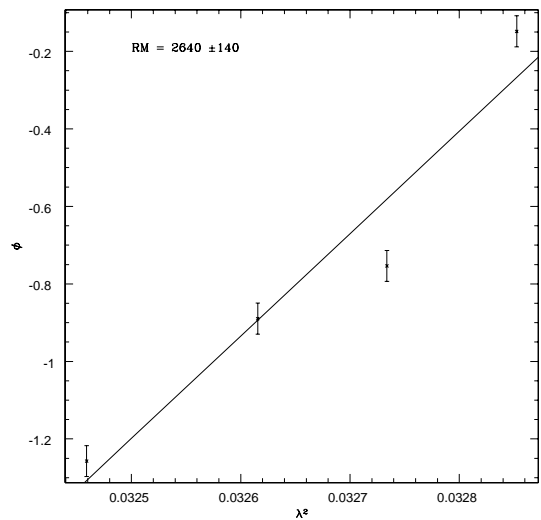}
   \caption{GRS1915+105 Faraday rotation in March 6 observation, wavelength squared is in metre squared, polarization angle in radians. intrinsic PA = -87 rad
            \label{fig:march_rm}
         }
   \end{figure}
%

   \begin{figure}
   \centering
   \includegraphics[angle=-90, width=7.5cm]{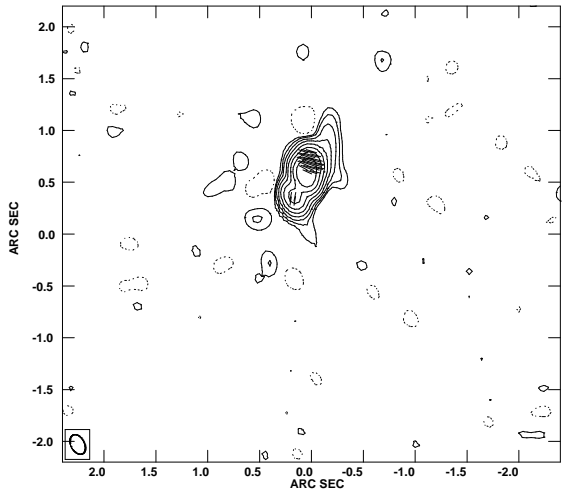}
   \caption{GRS1915+105 April 18 epoch, observed PA at 1658; MHz northen component $73.7^{o}$, southern component $-10.8^{o}$ .
            \label{fig:april}
         }
   \end{figure}
%

   \begin{figure}
   \centering
   \includegraphics[angle=-90, width=7.5cm]{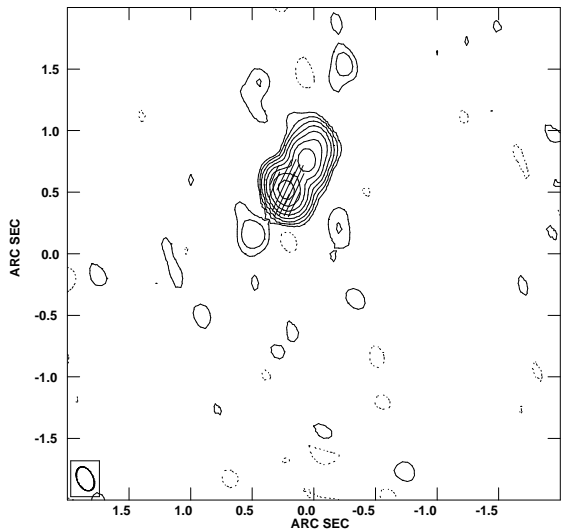}
      \caption{GRS1915+105 June 15. 13 \% polarized. Observed polarization angle at 1658 Mhz is $-24.9^{o}$.  }
         \label{fig:june}
   \end{figure}

   \begin{figure}
   \centering
   \includegraphics[width=7.5cm]{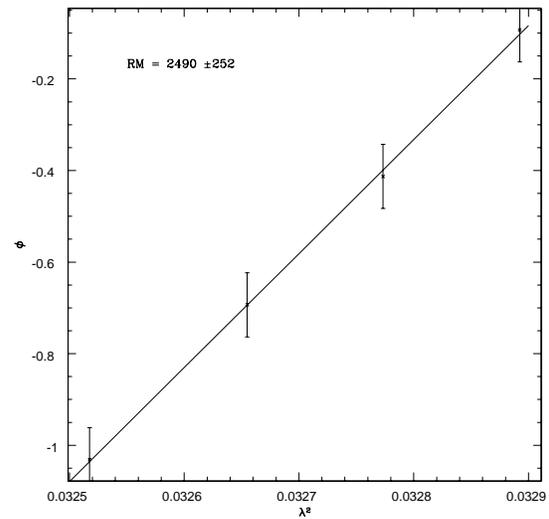}
   \caption{GRS1915+105 Faraday rotation in June 15-16 observation. intrinsic PA = -82 rad.  
            \label{fig:june_rm}
         }
   \end{figure}
%

\section{Results}

In  three images shown the  contours go  up in  powers of  two  from 0.24
mJy/beam. Due to 
instrumental problems no polarization could be detected for the March 24-25 
and May 9-10 observations. The 
Fig.~\ref{fig:march} image is  sightly extended to the north west and has polarization  
$\sim45^{o}$ to the  jet angle. The rotation measure in this epoch is 2640 
$\pm$ 140 rad/$m^2$.  In Fig.~\ref{fig:april} there is  polarization in the
northern  component  roughly  perpendicular   to  the  jet  axis  then
polarization   parallel  to  the   jet  axis   in  the   the  southern
component.  Neither  component shows  significant Faraday  rotation. Fig.~\ref{fig:june} shows  polarization only in  one component  and then
along the axis  of the jet. There is a rotation measure of 2490 $\pm$
252 rad/$m^2$ in this component.

A comparison can be made with two pulsars close to the line of sight,
PSR B1913+10 and PSR B1914+09. Their Faraday rotations were measured by
Hamilton \& Lyne (1987) to be 244.5 $\pm$ 1.0 and 61.1  $\pm$ 0.2 rad/$m^2$ respectively.
This suggests that the majority of the rotation is occurring inside the jet.

\section{Conclusions}

GRS 1915+105 shows in two epochs rotation measures of $\sim$ 2500 rad/$m^2$. looking at the rotation in pulsars close to the line of sight suggests that this is mainly occruing inside the jet. These results can be compared with the Faraday rotation seen in extragalactic jets ( e.g. Cotton et al. \cite{Cotton}). 

\begin{acknowledgements}

IKB acknowledges the receipt of a PPARC studentship and thanks Dr R.~E. Spencer for his help. MERLIN is operated by the University of Manchester as a national UK facility of PPARC.

\end{acknowledgements}

\end{document}